\documentclass[fleqn, usenatbib]{mnras}
\usepackage{amssymb,amsmath}
\usepackage{graphicx}
\usepackage{xcolor,natbib}
\usepackage{multirow}
\usepackage[T1]{fontenc}
\usepackage{ae,aecompl}
\usepackage{newtxtext, newtxmath}
\usepackage{float}
\usepackage{subfigure}
\usepackage{longtable}
\usepackage{enumitem}
\usepackage{longtable,listings}
\usepackage[flushleft]{threeparttable}
\usepackage{parcolumns}
\def\oc3{[O~{\sc iii}]$_c$}
\def\ob3{[O~{\sc iii}]$_b$}

\def\o3{[O~{\sc iii}]}

\title[NLRs size of SDSS J0838]
{Estimating sizes of Type 2 AGN narrow-line regions from multiple survey spectra - a demonstration}

\author[Zhang \& Zhao]
{Xue-Guang Zhang\thanks{Corresponding author Email: \href{mailto:xgzhang@njnu.edu.cn}{xgzhang@njnu.edu.cn}}, YuanBo Zhao\\
School of physics and technology, Nanjing Normal University, No. 1, Wenyuan Road, 210046, P. R. China
}

\pubyear{2022}

\begin{document}

\label{firstpage}

\pagerange{\pageref{firstpage}--\pageref{lastpage}}

\maketitle

\begin{abstract} 
	In the Letter, an interesting method is proposed to estimate size of narrow emission lines regions 
(NLRs) of a Type-2 AGN SDSS J083823.91+490241.1 (=SDSS J0838) at a redshift of 0.101, by comparing 
spectroscopic properties through the SDSS fiber (MJD=51873) (diameter of 3 arcseconds) and through the eBOSS 
fiber (MJD=55277) (diameter of 2 arcseconds). After subtractions of pPXF method determined host galaxy contributions, 
the narrow emission lines of SDSS J0838 in the SDSS spectrum and in the eBOSS spectrum can be well measured by 
Gaussian functions, leading more than 90\% of [O~{\sc iii}] emissions to be covered by the eBOSS fiber with 
diameter of 2 arcseconds. Meanwhile, both none broad emission components and none-variabilities of ZTF 3years-long 
g/r-band light curves can be applied to confirm SDSS J0838 as a Type-2 AGN, indicating few orientation effects 
on the projected NLRs size in SDSS J0838. Therefore, upper limit about 1arcsecond (2250pc) of the NLRs size 
can be reasonably accepted in SDSS J0838. Combining with the intrinsic reddening corrected [O~{\sc iii}] line 
luminosity, the upper limit of NLRs size in SDSS J0838 well lies within the 99.9999\% confidence bands of the 
R-L empirical relation for NLRs in AGN. 
\end{abstract}


\section{Introduction}

	Apparent narrow emission lines from narrow emission line regions (NLRs) are fundamental 
characteristics of Active Galactic Nuclei (AGN). Basic physical NLRs sizes (distance between NLRs and 
central power source) in AGN (especially in Type-2 narrow line AGN) have been well known, through 
properties of spatially resolved \o3~ emission images. \citet{bg06a, bg06b} have measured NLRs sizes  
in Seyfert 2 galaxy NGC1386 and the other several Seyfert-2 galaxies. \citet{gh11} have measured NLRs 
sizes of 15 obscured QSOs. \citet{liu13, lu13b} have measured NLRs sizes in a sample of radio quiet 
type-2 QSOs with extended emission-line regions. \citet{ha16} have measured 
NLR sizes in a nearby merging galaxy, to explore the dependence of NLR sizes on [O~{\sc iii}] luminosity 
in progenitors of dual AGN. \citet{fk18} have reported NLRs sizes in a sample of obscured QSOs. 
Based on the reported NLRs sizes in AGN, there is a strong dependence of NLRs size ($R_{NLRs}$) 
on [O~{\sc iii}]$\lambda5007$\AA~ line luminosity ($L_{[O~\textsc{iii}]}$) (the R-L empirical relation 
for NLRs) discussed in \citet{liu13, ha13, ha14, ha16, fk18, dz18},
\begin{equation}
\log(\frac{R_{NLRs}}{\rm pc})=(0.25\pm0.02)\times\log(\frac{L_{[O~\textsc{iii}]}}{\rm 10^{42}erg/s})
	 + (3.75\pm0.03)
\end{equation},
especially for AGN with 8$\mu$m luminosity smaller than $10^{45}{\rm erg/s}$ as discussed in \citet{ha13, 
dz18}.

	Until now, there are only around 50 AGN with NLRs sizes well measured by spatially resolved 
\o3~ emission images, more AGN with NLRs size being measured and/or estimated could provide further 
clues on intrinsic physical properties of NLRs of AGN. There are two probable sources of emission 
clouds in NLRs of AGN, local gas clouds related to star-formations (such as the emission regions in 
HII galaxies) and/or the ionized gas clouds related to central galactic outflows (as discussed in 
\citet{wy18}). Whether are there distinct locations for the two gas cloud sources is still unclear 
in AGN. In order to answer the question, it is necessary to find more AGN and HII galaxies of 
which NLRs sizes should be estimated. However, for AGN with weak [O~{\sc iii}] emissions and for HII 
galaxies with intrinsic weak [O~{\sc iii}] emissions, it would be difficult to determine NLRs 
([O~{\sc iii}] emission regions) sizes through spatially resolved high quality [O~{\sc iii}] emission 
images. Therefore, in the Letter, an interesting method proposed in Section 2 is applied to simply but 
conveniently check NLRs sizes properties of narrow emission line objects, based on the different 
fiber diameters of SDSS (Sloan Digital Sky Survey) and eBOSS (Extended Baryon Oscillation Spectroscopic 
Survey). Main results and discussions are shown in Section 3 on SDSS J083823.91+490241.1 
(=SDSS J0838) observed by both SDSS and eBOSS. Section 4 shows the final summaries and conclusions. 
And in the Letter, we have adopted the cosmological parameters of $H_{0}=70{\rm km\cdot s}^{-1}{\rm Mpc}^{-1}$, 
$\Omega_{\Lambda}=0.7$ and $\Omega_{\rm m}=0.3$.

\begin{figure*}
\centering\includegraphics[width = 18cm,height=3.5cm]{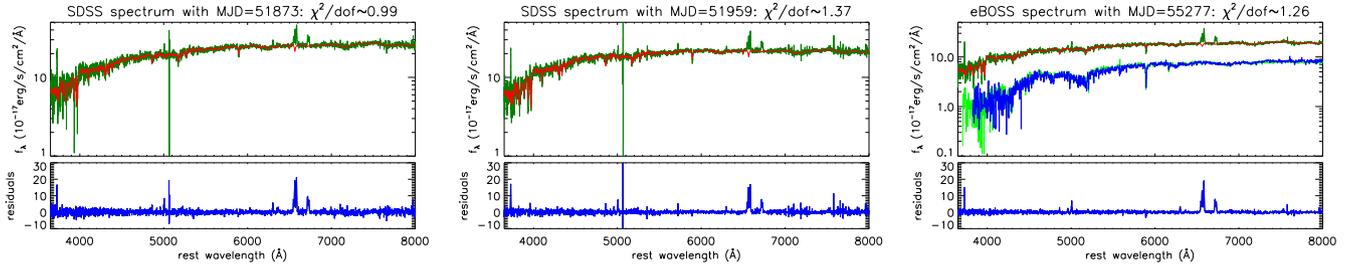}
\caption{Top panels show the SDSS spectra (left and middle panels) and the eBOSS spectrum (right 
panel) in dark green lines and the pPXF method determined host galaxy contributions in solid 
red lines in each panel. Each bottom panel shows the residuals calculated by the spectrum minus the 
pPXF method determined host galaxy contributions. In top right panel, solid line in blue 
and in green show the SDSS spectrum and the eBOSS spectrum of the collected K-5 star SDSS J083626.48+491230.9 
near to SDSS J0838.
}
\label{ssp}
\end{figure*}

\begin{figure*}
\centering\includegraphics[width = 18cm,height=7cm]{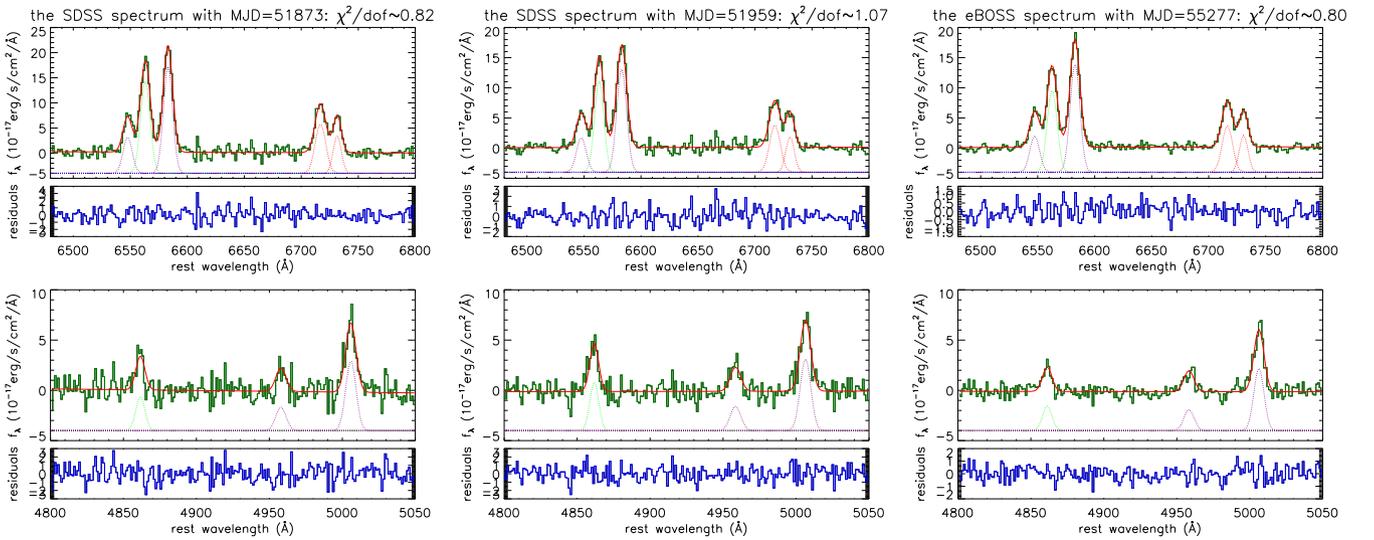}
\caption{Top panels show the line spectra around H$\alpha$ within rest wavelength from 6480\AA~ to 6800\AA~ 
in solid dark green lines by the SDSS spectra minus the host galaxy contributions (in left two panels) 
and by the eBOSS spectrum minus the host galaxy contributions (in right panels), and the best fitting results 
and the corresponding residuals to the emission lines shown as solid red lines by multiple Gaussian functions. 
In each top panel, dotted green line shows the determined narrow H$\alpha$, dotted purple lines show the 
determined [N~{\sc ii}] doublet, dotted red lines show the determined [S~{\sc ii}] doublet. Bottom panels 
show the results for the emission lines around H$\beta$ with rest wavelength range from 4800\AA~ to 5050\AA. 
In each bottom panel, dotted green line shows the determined narrow H$\beta$, dotted purple lines show the 
determined [O~{\sc iii}] doublet. The residuals are calculated by the line spectrum minus the best-fitting 
results. 
}
\label{line}
\end{figure*}

\begin{figure}
\centering\includegraphics[width = 6cm,height=4.5cm]{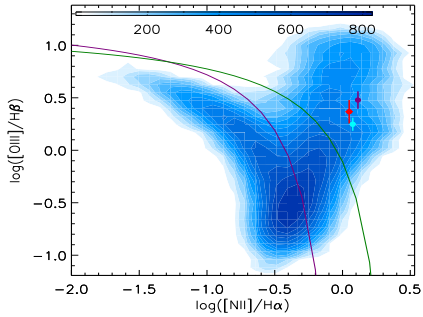}
\caption{BPT diagram of O3HB versus N2HA. Contour represents the results for all the narrow emission-line 
main galaxies collected from SDSS DR16. Solid circles in red, in cyan and in purple show the positions of 
SDSS J0838, relative to the emission line properties in the SDSS spectrum with MJD=51873, and in the SDSS 
spectrum with MJD=51959, and in the eBOSS spectrum with MJD=55277. Solid purple line and solid dark green 
line show the dividing lines reported in \citet{ka03a, ke06} between different kinds of narrow emission 
line objects.
}
\label{bpt}
\end{figure}

\section{Method}

	As the detailed descriptions in \url{https://www.sdss.org/dr16/spectro/spectro_basics/}, 
a spectrum in SDSS is observed by a fiber with diameter of 3 arcseconds, however, a spectrum in eBOSS 
is observed by a fiber with diameter of 2 arcseconds. If one object is observed by SDSS, and repeated 
by eBOSS, leading the multi-epoch spectra including different emission regions to have different 
expected properties of NLRs of the object.

	On the one hand, if emission lines in spectrum observed by eBOSS fiber have similar line 
intensities as the lines in spectrum by SDSS fiber, we would expect that the projected NLRs 
size of the object should be smaller than the eBOSS fiber radius (1 arcsecond). On the other hand, 
if emission lines in spectrum observed by eBOSS fiber have quite smaller line intensities than the 
lines in spectrum by SDSS fiber, we would expect that the projected size of NLRs of the object 
should be larger than the eBOSS fiber radius (1arcsecond).


\begin{table*}
\caption{Emission line parameters of SDSS J0838}
\begin{tabular}{cccccccccc}
\hline\hline
 & \multicolumn{3}{c}{in the SDSS spectrum with MJD=51873} & 
	\multicolumn{3}{c}{in the SDSS spectrum with MJD=51959} & 
	\multicolumn{3}{c}{in the eBOSS spectrum with MJD=55277} \\
 & \multicolumn{3}{c}{$\chi^2\sim0.82$} &
	\multicolumn{3}{c}{$\chi^2\sim1.07$} &
	\multicolumn{3}{c}{$\chi^2\sim0.80$}\\
\hline
	& $\lambda_0$ & $\sigma$ & flux & $\lambda_0$ & $\sigma$ & flux & $\lambda_0$ & $\sigma$ & flux\\
\hline
	H$\beta$  & 4861.63$\pm$0.11  & 3.09$\pm$0.11  & 26.2$\pm$3.9 & 
	4861.72$\pm$0.09 & 3.01$\pm$0.09 & 36.5$\pm$3.2 & 
	4861.21$\pm$0.08 & 3.22$\pm$0.09 & 19.6$\pm$2.3 \\
	O3 & 5005.86$\pm$0.34 & 3.52$\pm$0.35 & 61.1$\pm$5.8 &
	5005.75$\pm$0.78 & 4.75$\pm$0.78 & 64.8$\pm$3.6 & 
	5006.26$\pm$0.71 & 4.15$\pm$0.43 & 58.9$\pm$4.2 \\
	N2 & 6583.09$\pm$0.14 & 4.03$\pm$0.14 & 214.6$\pm$6.9 & 
	6583.02$\pm$0.12 & 4.28$\pm$0.12 & 184.1$\pm$4.8 & 
	6582.72$\pm$0.09 & 4.26$\pm$0.09 & 192.5$\pm$3.5 \\
	H$\alpha$ & 6563.19$\pm$0.14 & 4.18$\pm$0.14 & 191.4$\pm$6.1 & 
	6563.31$\pm$0.12 & 4.06$\pm$0.12 & 154.7$\pm$4.2 &
	6562.61$\pm$0.11 & 4.34$\pm$0.12 & 148.2$\pm$3.6 \\
	S16 & 6716.85$\pm$0.35 & 4.65$\pm$0.37 & 114.2$\pm$7.6 & 
	6717.46$\pm$0.37 & 4.73$\pm$0.37 & 92.4$\pm$6.4 & 
	6716.35$\pm$0.25 & 4.61$\pm$0.26 & 88.6$\pm$4.4 \\
	S31 &  6731.55$\pm$0.38 & 3.54$\pm$0.39 & 65.9$\pm$6.5 & 
	6730.74$\pm$0.41 & 3.69$\pm$0.39 & 54.7$\pm$5.7 &
	6730.62$\pm$0.28 & 3.81$\pm$0.28 & 58.1$\pm$3.9 \\
\hline
\end{tabular}\\
Notice: $\lambda_0$, $\sigma$ and flux represent central wavelength in unit of \AA, second moment in 
unit of \AA, and line intensity in unit of $10^{-17}{\rm erg/s/cm^2}$. "O3" means the 
[O~{\sc iii}]$\lambda5007$\AA~ emission line. "N2" means the [N~{\sc ii}]$\lambda6583$\AA~ emission 
line. "S16" and "S31" mean the [S~{\sc ii}]$\lambda6716$\AA~ emission line and the [S~{\sc ii}]$\lambda6731$\AA~ 
emission line. $\chi^2$ are calculated by the summed squared residuals divided by the degree of freedom.
\end{table*}

\section{Main Results} 

\subsection{spectroscopic results}

	Fig.~\ref{ssp} shows the spectra of SDSS J0838 at a redshift 0.101 with plate-mjd-fiberid=0445-51873-0124 
and with plate-mjd-fiberid=0550-51959-0411 (the SDSS spectra, with drilled fiber position of plug\_ra=129.59965 
and plug\_dec=49.044810) and with plate-mjd-fiberid=3696-55277-0030 (the eBOSS spectrum, with drilled fiber 
position of plug\_ra=129.59963 and plug\_dec= 49.044865), collected from SDSS DR16 (data release 16, \citet{ap20}). 
It is clear that the SDSS fiber and the eBOSS fiber have been positioned to the same central position with a 
difference of about 0.2arcseconds, and the SDSS spectrum and the eBOSS spectrum are quite different, due to 
different peak intensities.

	Actually, among low redshift ($z<0.3$) AGN in SDSS DR16, hundreds of AGN have been collected 
to have both SDSS spectra and eBOSS spectra. However, among the objects discussed in detail in our manuscripts 
in preparation, SDSS J0838 is selected as the main target in the Letter by the following three main reasons. 
First, SDSS J0838 is a well classified Type-2 AGN discussed in the following sections, indicating few orientation 
effects on estimated NLRs sizes in SDSS 0838. Second, SDSS 0838 has its SDSS spectrum and eBOSS spectrum with 
almost totally the same drilled fiber positions, leading to few effects of pointing error on estimated NLRs 
sizes in SDSS 0838. Third, SDSS J0838 has its single-Gaussian described [O~{\sc iii}] emissions almost totally 
covered by the eBOSS fiber leading to clear upper limits of NLRs sizes.

	Before proceeding further, the spectroscopic features of the SDSS spectra and the eBOSS spectrum 
are carefully checked in SDSS J0838, in order to measure the emission lines after subtractions of 
starlight. In the Letter, the pPXF (Penalized Pixel-Fitting) code \citep{ref32}, one commonly 
applied SSP (Simple Stellar Population) method \citep{ref33, kh03, cp13, wc19, zh19}, is accepted to determine 
contributions of starlight, applied with 224 SSP templates collected from the MILES (Medium 
resolution INT Library of Empirical Spectra) stellar library \citep{ref34, ks21} with 32 stellar ages from 
0.06Gyrs to 17.78Gyrs and with 7 metallicities from -2.32 to 0.22. After considering the popular regularization 
method, the pPXF method can give reliable and smoother star-formation histories. Fig.~\ref{ssp} shows the pPXF 
method determined starlight in the spectra of SDSS J0838.

	After subtractions of the pPXF method determined starlight, emission lines can be well 
measured. Here, the narrow emission lines of H$\alpha$, H$\beta$, [O~{\sc iii}]$\lambda4959, 5007$\AA~ doublet, 
[N~{\sc ii}]$\lambda6548, 6583$\AA~ doublet and [S~{\sc ii}]$\lambda6716,6731$\AA~ doublet are mainly considered. 
One Gaussian component is accepted to describe each narrow emission line. Due to few effects of broad emission 
lines, there are no server restrictions on the model parameters, besides the following four simple restrictions. 
First, the flux ratio of [O~{\sc iii}] ([N~{\sc ii}]) doublet is fixed to the theoretical value of 
2.99 \citep{sz00, dp07} (3.05 \citep{dp22}). Second, the components of H$\alpha$ and H$\beta$ have the same 
redshift and the same line width (in velocity space). Third, the components of each forbidden line doublet 
have the same redshift and the same line width. Fourth, emission intensity of each Gaussian component is not 
smaller than zero. Through the Levenberg-Marquardt least-squares minimization technique (the MPFIT package), 
the best fitting results and the corresponding residuals (line spectrum minus the best fitting results) 
to the emission lines are shown in Fig.~\ref{line}. The measured line parameters are listed in Table~1. 
Here, as the shown besting fitting results to the [O~{\sc iii}] doublets in bottom panels of Fig.~\ref{line}, 
it is not necessary to consider extended components in [O~{\sc iii}] doublets as discussed in \citet{ref35, 
ref36, ref37} for [O~{\sc iii}] doublet in SDSS J0838.

	Based on the measured line parameters, the following three points can be confirmed. First and foremost, 
more than 90\% of the [O~{\sc iii}] emissions are coming from the eBOSS fiber covered emission regions. 
Considering the SDSS fiber covered area about 2.25 times larger than the area covered by the eBOSS fiber, 
we can safely accept that the projected distance of [O~{\sc iii}] emission regions should be smaller than 
the eBOSS fiber radius 1arcsecond (about 2250pc) in SDSS J0838.

	Besides, flux ratios of O3HB ([O~{\sc iii}]$\lambda5007$\AA~ to narrow H$\beta$) and N2HA 
([N~{\sc ii}]$\lambda6583$\AA~ to narrow H$\alpha$) are about $2.33_{-0.49}^{+0.67}$ and $1.12_{-0.07}^{+0.07}$ 
for the lines in the SDSS spectrum with MJD=51873, about $1.77_{-0.23}^{+0.28}$ and $1.19_{-0.06}^{+0.07}$ 
for the lines in the SDSS spectrum with MJD=51959, about $3.01_{-0.51}^{+0.63}$ and $1.30_{-0.05}^{+0.06}$ 
for the lines in the eBOSS spectrum with MJD=55277, leading SDSS J0838 to be classified as an AGN based 
on the dividing lines in the BPT diagram \citep{bpt81, ka03a, ke06, km13, ref20, ref21, ref22, zh22} shown 
in Fig.~\ref{bpt}. Moreover, based on the best fitting results to the emission lines shown in Fig.~\ref{line}, 
there are no clues for broad emission lines, indicating SDSS J0838 is a Type-2 AGN. Moreover, from inner 
region covered in the eBOSS fiber to outer region covered in the SDSS fiber of the [O~{\sc iii}] emissions, 
the lower O3HB at longer distance is similar as the trends of O3HB on radius shown in Fig.~6\ in \citet{liu13} 
for obscured radio quiet quasars.

	Last but not the least, Balmer decrements (flux ratio of narrow H$\alpha$ to narrow H$\beta$) are 
about $7.3_{-1.1}^{+1.6}$, about $4.2_{-0.45}^{+0.53}$, about $7.6_{-0.9}^{+1.2}$, based on the Balmer lines 
in the SDSS spectrum with MJD=51873, in the SDSS spectrum with MJD=51959, and in the eBOSS spectrum with MJD=55277, 
respectively. It is strange that there are different Balmer decrements in narrow emission line regions, between the 
double observations by the SDSS fiber with MJD=51873 and with MJD=51959, unless there were moving dust clouds. 
However, if considering corrections of reddening effects on the [O~{\sc iii}] emissions after accepted the intrinsic 
flux ratio of narrow H$\alpha$ to narrow H$\beta$ to be 3.1, intrinsic [O~{\sc iii}] emissions in the SDSS spectrum 
with MJD=51959 (fiber radius 1.5arcseconds) with $E(B-V)=0.26$ would be $149\times10^{-17}{\rm erg/s/cm^2}$, 
apparently smaller than the expected intrinsic [O~{\sc iii}] emissions about $690\times10^{-17}{\rm erg/s/cm^2}$ 
in the eBOSS spectrum with MJD=55277 (fiber radius 1arcsecond) with $E(B-V)=0.77$. The unreasonable result indicates 
the results from the SDSS spectrum with MJD=51959 cannot be accepted. Therefore, the measured line parameters in the 
SDSS spectrum with MJD=51959 are not considered nor discussed further in the Letter, besides listed in Table~1. 
Unfortunately, we do not have a reasonable explanation to the strange Balmer decrements in NLRs in the SDSS spectrum 
with MJD=51959. 

\begin{figure}
\centering\includegraphics[width = 8cm,height=4.75cm]{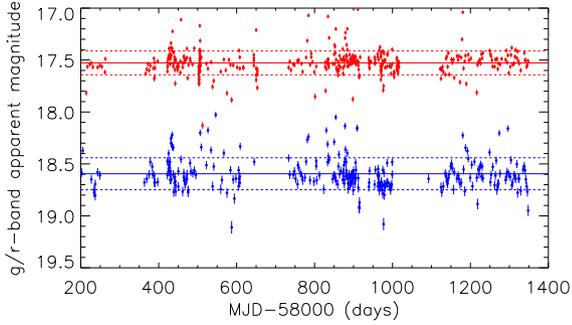}
\caption{ZTF g/r-band long-term variabilities of SDSS J0838. Solid circles in blue and in red show the g-band 
and the r-band light curves, respectively. Horizontal blue and red lines show positions of mean magnitudes of 
the g-band and r-band light curves. Horizontal dashed blue line and dashed red lines show bands relative to 
standard deviations of the g-band and the r-band light curves, respectively.
}
\label{lmc}
\end{figure}

\begin{figure}
\centering\includegraphics[width = 8cm,height=4.75cm]{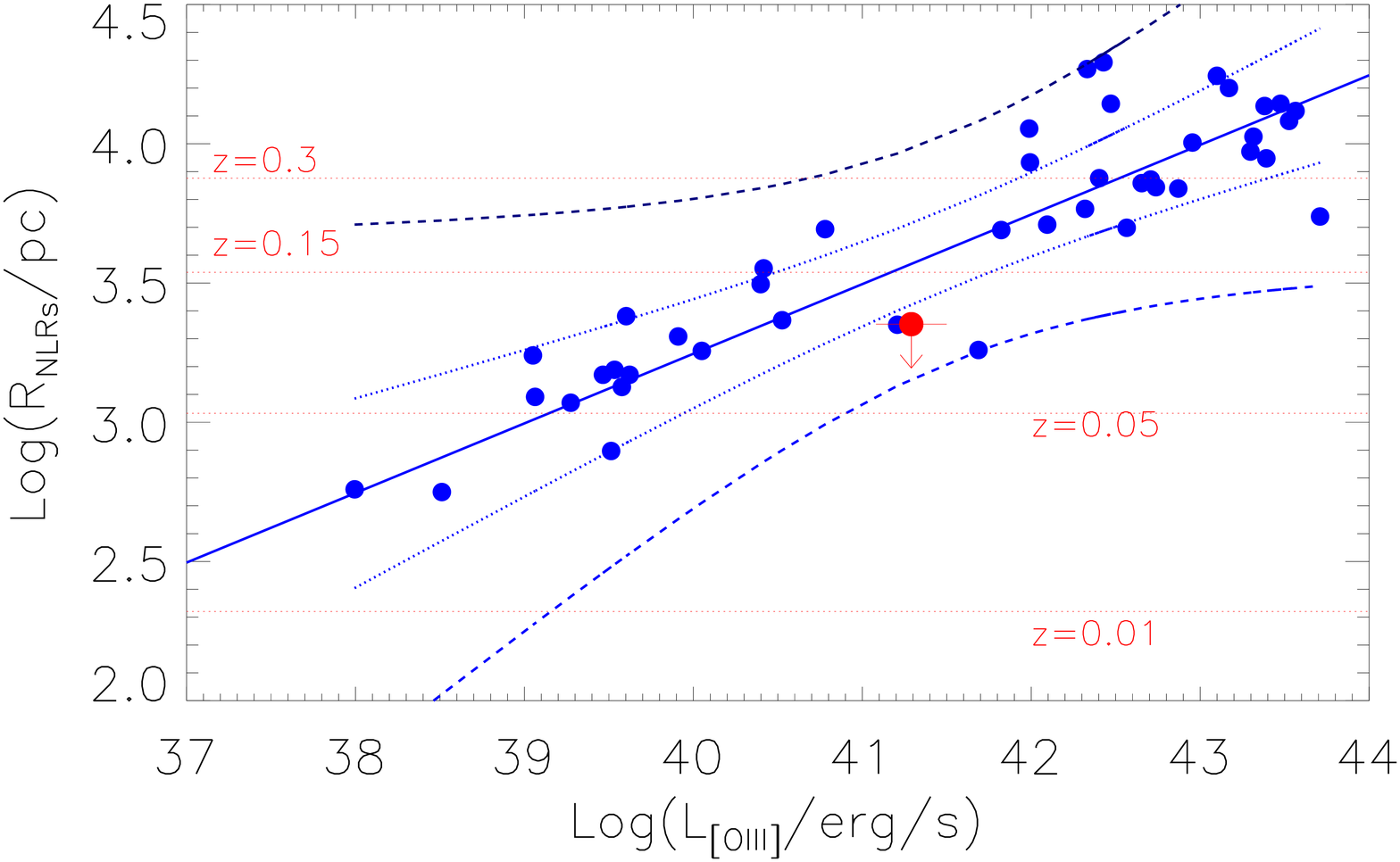}
\caption{Dependence of NLRs size on [O~{\sc iii}] line luminosity of AGN. Solid circles are the data points 
collected from \citet{liu13}, solid blue line shows the best fitting results (Equation 1\ in the Introduction), 
dotted and dashed blue lines show the 90\% and 99.9999\% confidence bands to the best-fitting results. Solid 
red circle shows the results in SDSS J0838 in the Letter. Horizontal dotted red lines mark the projected 
distance relative to 1arcsecond for different redshifts.}
\label{rl}
\end{figure}

	Before the end of the subsection, some further descriptions are given on flux calibrations in 
SDSS spectra which does affect both spectral slope and ratios of well-separated features (such as the 
Balmer decrement). More detailed descriptions on flux calibrations can be found in 
\url{https://www.sdss.org/dr16/algorithms/spectrophotometry/}. In SDSS DR16, flux calibration processes 
have been greatly improved, with a new set of stellar templates applied to fit absorption lines of standard 
stars, leading flux calibration residuals to be reduced by a factor of 2\ in the blue spectrograph (3600 
to 6000\AA). While, expected scatter for stars between $ugriz$ fluxes from the images and fluxes synthesized
from the calibrated spectra is 4\% for the original SDSS and rises to 6\% with the smaller BOSS fibers, 
strongly indicating the flux calibration uncertainties have few effects on our final results. Moreover, 
by an amount depending on airmass and position on plates, spectra of classified quasars in eBOSS are 
biased blue with a typical spectral-index offset of 0.4, leading to about 12\% change in Balmer decrement 
in quasars in eBOSS. However, SDSS J0838 is classified as GALAXY, indicating the airmass and position on 
plates have few effects on our results. Furthermore, as well discussed in \citet{fa82}, atmospheric 
differential refractions have important effects on flux calibrations. However, near to SDSS J0838, there 
is a K-5 star (SDSS J083626.48+491230.9) collected with both SDSS spectrum (plate-mjd-fiberid=0443-51873-0609) 
and eBOSS spectrum (plate-mjd-fiberid=3697-55290-0284) shown in top right panel of Fig.~\ref{ssp} with 
continuum intensity difference about 5\% at 5000\AA. Therefore, effects of atmospheric differential 
refractions have been totally considered in flux calibrations.

\subsection{Photometric Variabilities}

	Based on the spectroscopic properties, there are apparent narrow emission lines but no broad Balmer 
lines. The SDSS J0838 can be identified as a Type-2 AGN. In the subsection, long-term photometric variabilities 
are applied to re-confirm SDSS J0838 as a Type-2 AGN. The 3years-long g/r-band light curves can be collected 
from the ZTF (Zwicky Transient Facility) \citep{be19, gk19, ds20} from Mar. 2018 to Aug. 2021, and shown in 
Fig.~\ref{lmc}, well applied to re-confirm that the SDSS J0838 is a Type-2 AGN, not a true Type-2 AGN without 
hidden central broad emission line regions. More discussions on properties of true type-2 AGN with apparent 
variabilities can be found in \citet{bv14, ly15, pw16, zh21c}.  

\subsection{NLRs size of SDSS J0838}

	Once SDSS J0838 is well identified as a Type-2 AGN, there are few orientation effects on 
the NLRs size (the distance between [O~{\sc iii}] emission regions and central black hole) in SDSS J0838. 
Therefore, the 2250pc (1arcsecond) can be well accepted as the upper limit of NLRs size in SDSS J0838. 
Considering the reddening corrections with $E(B-V)\sim0.76\pm0.15$ applied, the intrinsic 
[O~{\sc iii}]$\lambda5007$\AA~ line luminosity can be estimated as 
$L_{[O~\textsc{iii}]}\sim(1.95\pm0.94)\times10^{41}{\rm erg/s}$, with the uncertainty 
of $L_{[O~\textsc{iii}]}$ estimated by uncertainty of the applied E(B-V). Then, we can check the properties 
of NLRs size of SDSS J0838 through the R-L empirical relation for NLRs in AGN, shown in Fig.~\ref{rl}. 
Considering the 90\% and 99.9999\% confidence bands to the R-L empirical relation for the NLRs in AGN shown 
in Fig.~\ref{rl}, we can safely conclude the upper limits of NLRs size in SDSS J0838 is within the 99.9999\% 
confidence interval of the expected results from the empirical R-L relation for NLRs in type-2 AGN.

	In the near future, samples of Type-2 AGN and HII galaxies with redshift smaller than 0.35 will 
be discussed in detailed in our manuscripts in preparation, through the spectra observed by 
both the SDSS fibers and the eBOSS fibers. As the shown projected distances in Fig.~\ref{rl} for the 1arcsecond (eBOSS 
fiber radius) relative to different redshifts, at least three interesting results will be mainly focused 
on. First, there would probably be some special Type-2 AGN with redshift around 0.3 (with redshift around 
0.05) but lower [O~{\sc iii}] line luminosity (stronger 
[O~{\sc iii}] line luminosity), lying outside of the 99.9999\% confidence interval of the empirical R-L 
relation for NLRs in the type-2 AGN, to provide unique candidates of Type-2 AGN with probably special NLRs 
properties not consistent with the R-L empirical relation expected results. Second, for common 
Type-2 AGN and HII galaxies, it will be interesting to check are there different properties in the space 
of NLRs size versus [O~{\sc iii}] line luminosity, to provide further clues on origin of gas clouds in 
NLRs of AGN. Third, it will lead to more detected luminous extended narrow emission line regions 
(sizes larger 10kpcs) which are so rare that any systematic ways to identify them are useful even if many 
false positives remain.

\section{Summaries and Conclusions}

The final summaries and conclusions are as follows.
\begin{itemize}
\item Due to different fiber diameters between the SDSS fibers and the eBOSS fibers, upper limits of NLRs 
(especially the [O~{\sc iii}] emission regions) sizes of Type-2 AGN with few effects of orientations 
could be simply estimated, once the [O~{\sc iii}] emissions are totally covered into the eBOSS fiber, 
based on similar [O~{\sc iii}] emission intensities in the SDSS spectrum and in the eBOSS spectrum.
\item Based on the besting-fitting results to the emission lines after subtractions of pPXF method determined 
host galaxy contributions in SDSS J0838, there are no apparent broad Balmer emission lines in SDSS J0838, 
indicating SDSS J0838 is a Type-2 AGN.
\item Emission lines can be well measured in the Type-2 AGN SDSS J0838 with spectra observed through both the 
SDSS fiber and the eBOSS fiber. And more than 90\% of [O~{\sc iii}] emissions are covered into the eBOSS fiber, 
indicating the eBOSS fiber radius can be accepted as the upper limit of NLRs size in SDSS J0838.
\item Based on the ZTF 3years-long g/r-band light curves, there are none variabilities in SDSS J0838, to 
re-confirm SDSS J0838 as a Type-2 AGN, indicating few effects of orientation on the projected NLRs size in SDSS 
J0838.
\item Based on the intrinsic reddening corrected [O~{\sc iii}] line luminosity and the upper limit of NLRs 
size of SDSS J0838, the upper limit of NLRs size in SDSS J0838 is within the 99.9999\% confidence interval 
of the expected results from the empirical R-L relation for NLRs in type-2 AGN.
\end{itemize}

\section*{Acknowledgements}
Zhang gratefully acknowledge the anonymous referee for giving us constructive comments
and suggestions to greatly improve the paper.
Zhang gratefully thanks the kind grant support from NSFC-12173020. This manuscript has made use of the data 
from the SDSS projects, \url{http://www.sdss3.org/}, managed by the Astrophysical Research Consortium 
for the Participating Institutions of the SDSS-III Collaborations. This 
manuscript has made use of the data from the ZTF \url{https://www.ztf.caltech.edu}, and made use of the MPFIT 
package \url{https://pages.physics.wisc.edu/~craigm/idl/cmpfit.html}, and of the pPXF code 
\url{http://www-astro.physics.ox.ac.uk/~mxc/software/#ppxf}.

\section*{Data Availability}
The data underlying this article will be shared on reasonable request to the corresponding author
(\href{mailto:xgzhang@njnu.edu.cn}{xgzhang@njnu.edu.cn}).

\label{lastpage}
\end{document}